\documentclass{iopart}
\usepackage{iopams}

\usepackage{hyperref}
\usepackage{graphicx}
\usepackage{mathptmx}
\usepackage[ngerman,american]{babel}
\usepackage{cite}
\usepackage{url}
\usepackage{textcomp}
\usepackage{wasysym}

\usepackage[dvipsnames,svgnames]{pstricks}
\usepackage{pstricks-add}
\usepackage{pst-3d}
\usepackage{pst-3dplot}
\usepackage{pst-all, pst-eucl,pst-plot,pst-math,pst-poly,pst-node}

\newcommand{\bigo}[1]{\mathcal{O}\left(#1\right)}

\newcommand{\DX}{\mathrm{DX}}
\newcommand{\pc}{p_{\mathrm{c}}}
\newcommand{\dc}{d_{\mathrm{c}}}
\newcommand{\smax}{n_{\mathrm{max}}}
\newcommand{\dmax}{d_{\mathrm{max}}}

\newcommand{\eqref}[1]{(\ref{#1})}

\newcommand{\cube}[1]{{\pstThreeDBox[fillstyle=solid,fillcolor=green,RotZ=0](#1)(0,0,1)(0,1,0)(1,0,0)}}
\newcommand{\cell}[1]{\pscustom[fillstyle=solid,fillcolor=green]{\translate(#1)
  \psframe(0,0)(1,1) \translate(0,0)}}
\newcommand{\graphnodes}{\psset{fillstyle=solid,fillcolor=white,radius=4pt,shadow=true,shadowcolor=black!50}}

\newcommand{\seqnum}[1]{\href{http://oeis.org/#1}{#1}}

\savedata{\crystalballdata}[
{{1, 0.}, {2, 1.27875}, {3, 2.25768}, {4, 3.06408}, {5, 3.75136}, {6, 
  4.34953}, {7, 4.87804}, {8, 5.35053}, {9, 5.777}, {10, 
  6.16511}, {11, 6.5208}, {12, 6.84879}}]
\savedata{\crystalcubedata}[
{{1, 0.}, {2, 4.29409}, {3, 6.29073}, {4, 7.60588}, {5, 8.58818}, {6, 
  9.37253}, {7, 10.0255}, {8, 10.5848}, {9, 11.074}, {10, 
  11.5088}, {11, 11.9}, {12, 12.2556}}]

\savedata{\lambdaninedata}[{
{0.0788157,3.67181},
{0.0591042,3.67759},
{0.0463379,3.68232},
{0.0375311,3.68565},
{0.0311632,3.68813},
{0.0263882,3.69002},
{0.0227022,3.69148},
{0.0197886,3.69265},
{0.0174397,3.69359},
{0.0155144,3.69436}
}]

\savedata{\thetaninedata}[{
{0.50235,	2.04401},
{0.449654,2.07814},
{0.41073,2.10405},
{0.38046,2.12998},
{0.356049,2.15158},
{0.33582,2.17018},
{0.3187,2.1862},
{0.303966,2.20013},
{0.291111,2.21234},
{0.279766,2.22314},
{0.269657,2.23277}
}]

\begin{document}

\title{Counting Lattice Animals in High Dimensions}

\author{Sebastian Luther$^1$ and Stephan Mertens$^{1,2}$}

\address{\selectlanguage{ngerman}{$^1$Institut\ f"ur\ Theoretische\ Physik,
    Otto-von-Guericke Universit"at, PF~4120, 39016 Magdeburg,
    Germany}} 

\address{$^2$Santa Fe Institute,
1399 Hyde Park Rd,
Santa Fe, NM 87501,
USA}

\ead{SebastianLuther@gmx.de, mertens@ovgu.de}

\begin{abstract}
  We present an implementation of Redelemeier's algorithm for the
  enumeration of lattice animals in high dimensional lattices. The
  implementation is lean and fast enough to allow us to
  extend the existing tables of animal counts, perimeter polynomials
  and series expansion coefficients in $d$-dimensional hypercubic
  lattices for $3 \leq d\leq 10$. From the data we compute
  formulas for perimeter polynomials for lattice animals of size
  $n\leq 11$ in arbitrary dimension $d$.  When amended by
  combinatorial arguments, the new data suffices to yield explicit
  formulas for the number of lattice animals of size $n\leq 14$ and
  arbitrary $d$.  We also use the enumeration data to compute
  numerical estimates for growth rates and exponents in high
  dimensions that agree very well with Monte Carlo simulations and
  recent predictions from field theory.
\end{abstract}




\section{Introduction}
\label{sec:intro}

A \emph{polyomino} of size $n$ is an edge-connected set of $n$ squares on the
square lattice, a \emph{polycube}  of size $n$ is a face-connected set of $n$ cubes in the
cubic lattice. Polyominoes and polycubes are a classical topic in recreational
mathematics and combinatorics \cite{golomb:book}. In statistical
physics and percolation theory, polyominoes and polycubes are called
\emph{lattice animals} \cite{stauffer:aharony:book,guttmann:LNP}.
Lattice animals are not restricted to dimension $2$ or $3$:
a $d$-dimensional lattice animal of size $n$ is a set of $n$
face-connected hypercubes on $\mathbb{Z}^d$.

\begin{figure}
\centering
\psset{unit=0.04\columnwidth}
\begin{pspicture}(0,-1.5)(17,3)
  \cell{0,0}\cell{0,1}\cell{1,1}
  \cell{3,1}\cell{4,1}\cell{4,0}
  \cell{6,0}\cell{7,0}\cell{7,1}
  \cell{9,1}\cell{9,0}\cell{10,0}
  \cell{12,0}\cell{12,1}\cell{12,2}
  \cell{14,0}\cell{15,0}\cell{16,0}
  \rput(1,-0.5){$t=7$}
  \rput(4,-0.5){$t=7$}
  \rput(7,-0.5){$t=7$}
  \rput(10,-0.5){$t=7$}
  \rput(12.5,-0.5){$t=8$}
  \rput(15.5,-0.5){$t=8$}
\end{pspicture}\\
\psset{unit=0.04\columnwidth}
\begin{pspicture}(-1,-2.5)(20.5,3)
\psset{RotZ=45}
\pstThreeDPut(0,0,0){ \cube{0,0,0} \cube{0,0,1}  \cube{0,0,2} \cube{0,1,0} }
\rput(0,-2){$t={17}$}
\pstThreeDPut(0,3,0){ \cube{0,0,0} \cube{0,0,1}  \cube{0,0,2} \cube{0,1,1} }
\rput(3,-2){$t={16}$}
\pstThreeDPut(0,6,0){ \cube{0,1,0} \cube{0,0,1}  \cube{0,0,2} \cube{0,1,1} }
\rput(6,-2){$t={16}$}
\pstThreeDPut(0,9,0){ \cube{0,0,-1} \cube{0,0,0} \cube{0,0,1}  \cube{0,0,2}  }
\rput(9,-2){$t={18}$}
\pstThreeDPut(0,12,0){ \cube{0,0,0} \cube{0,1,0}  \cube{1,0,0} \cube{1,1,0} }
\rput(12,-2){$t={16}$}
\pstThreeDPut(0,15.5,0){ \cube{0,0,0} \cube{0,1,0}  \cube{1,0,0} \cube{0,0,1} }
\rput(15.5,-2){$t={15}$}
\pstThreeDPut(0,19,0){  \cube{0,1,0}  \cube{1,0,0} \cube{1,1,0} \cube{0,1,1}}
\rput(19,-2){$t={16}$}
\end{pspicture}
\caption{All \emph{fixed} polyominoes of size $n=3$ (top) and all
  \emph{free} polycubes of size $n=4$ (bottom) and
  their perimeters.\label{fig:polycubes}}
\end{figure}
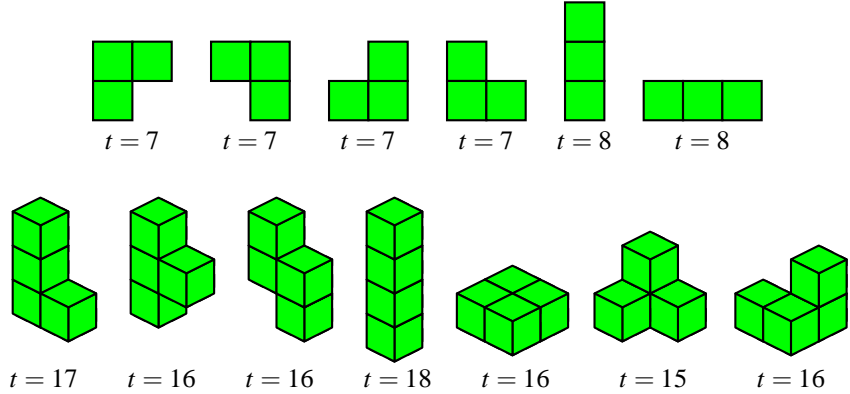

In this contribution we will address the problem of counting the
number of \emph{fixed} animals of size $n$ in dimension $d > 2$.
\emph{Fixed} animals are considered distinct if
they have different shapes or orientations. \emph{Free} animals, on
the other hand,
are distinguished only by shape, not by orientation. 
Figure~\ref{fig:polycubes} shows all fixed polyominoes of size $3$ and
all free polycubes of size $4$. 

We denote the number of $d$-dimensional fixed animals of
size $n$ by $A_d(n)$. There is no formula for $A_d(n)$, but we know that $A_d(n)$
grows exponentially with $n$. Using subadditivity and concatenation
arguments \cite{klarner:67}, one can show that there are constants $1
< \lambda_d < \infty$ such that
\begin{equation}
  \label{eq:klarner}
  \lim_{n\to\infty} \sqrt[n]{A_d(n)} = \lambda_d\,.
\end{equation}
The constant $\lambda_2$ is known as Klarner's constant.

A slightly stronger result due to Madras \cite{madras:99} asserts that
\begin{equation}
  \label{eq:madras}
  \lim_{n\to\infty} \frac{A_d(n+1)}{A_d(n)} = \lambda_d\,.
\end{equation}
Intuitively, the growth rate $\lambda_d$ should grow with the
coordination number $2d$ of the lattice. In fact, in
\cite{barequet:barequet:rote:10} it is shown that
\begin{equation}
  \label{eq:5}
  \lambda_d = 2d\mathrm{e} - o(d)\,,
\end{equation}
and in the same paper it is conjectured that $\lambda_d = (2d-3)\mathrm{e} + O(1/d)$. 
For finite $d$, however, we know only lower and upper bounds for $\lambda_d$. 
Numerical estimates for $\lambda_d$ can be derived from extrapolating
$A_d(n+1)/A_d(n)$, which is one motivation to compute $A_d(n)$ for $n$
as large as possible. We will try our hands at that in Section~\ref{sec:lambda}.

In percolation theory one is interested in counting lattice animals of
a given size $n$ according to their perimeter $t$, i.e., to the number
of adjacent cells that are empty (see Figure~\ref{fig:polycubes}).
If each cell of the lattice is occupied independently with probability $p$, the
average number of clusters of size $n$ per lattice site reads
\begin{equation}
  \label{eq:def-ns}
  \sum_{t} g_{n,t}^{(d)} p^n(1-p)^t\,,
\end{equation}
where $g_{n,t}^{(d)}$ denotes the number of fixed $d$-dimensional lattice
animals of size $n$ and perimeter $t$. The $g$'s define the 
\emph{perimeter polynomials} 
\begin{equation}
  \label{eq:perimeter-polynomial}
  P_d(n,q) = \sum_t g_{n,t}^{(d)} q^t\,. 
\end{equation}
We can easily compute $A_d(n)$ from the perimeter polynomial
$P_d(n,q)$ through
\begin{equation}
  \label{eq:A-from-P}
  A_d(n) = P_d(n,1) = \sum_t g_{n,t}^{(d)}\,.
\end{equation}
In fact we can compute $A_d(n+1)$ from the perimeter polynomials
up to size $n$,
\begin{equation}
  \label{eq:extra-A-from-P}
  A_d(n+1) = \frac{1}{n+1}\sum_{m \leq n} m\,\sum_t g_{m,t}^{(d)} 
{t\choose n+1-m}(-1)^{n-m}\,. 
\end{equation}
This equation follows from the observation that below the percolation
threshold $\pc$, each occupied lattice
site belongs to some finite size lattice animal,
\begin{equation}
  \label{eq:p-conservation}
  p = \sum_{n=1}^\infty n p^n P_d(n,1-p) 
\end{equation}
for $p < \pc$. The right hand side is a power series in
$p$, and Equation~\eqref{eq:extra-A-from-P} follows from the fact that the
coefficient of $p^{n+1}$ must be zero.

As we will see in the next section, the algorithm for counting lattice
animals keeps track of the perimeter anyway.  Hence it is reasonable
to use the algorithm to compute the perimeter polynomials and to apply
\eqref{eq:extra-A-from-P} to get an extra value of $A_d$.

\section{The Algorithm}
\label{sec:algorithm}

The classical algorithm for counting lattice animals is due to Redelmeier
\cite{redelmeier:81}. Originally developped for the square lattice,
Redelmeier's algorithm was later shown to work on arbitrary
lattices and in higher dimensions \cite{mertens:90} and to be
efficiently parallelizable \cite{mertens:lautenbacher:92}.
For two dimensional lattices there is a much faster counting method based
on transfer matrices \cite{jensen:01}, but for $d\geq 3$ Redelmeier's
algorithm is still the most efficient known way to count lattice animals. 

The algorithm works by recursively generating all lattice animals up
to a given size $\smax$. Given an animal of size $n$, the algorithm
generates animals of size $n+1$ by adding a new cell in the
perimeter of  the given animal. The lattice sites that are available
for extending the current animal are stored in a set $U$ called the
\emph{untried set}. To avoid generating the same fixed animal more
than once, lattice sites that have previously been added to the untried set are
marked on the lattice.

\begin{figure}[h!]
\centering
\psset{unit=0.065\columnwidth}
\begin{pspicture}(-1,-1)(6,4)
\psset{linecolor=black!25,fillstyle=solid,fillcolor=black!25}
\psframe(2,0)(3,1)
\psframe(2,1)(3,2)
\psframe(1,1)(2,2)
\psframe(1,2)(2,3)
\psset{linecolor=black,fillstyle=none}
\psline[linestyle=solid](-1,4)(6,4)
\psline[linestyle=solid](-1,3)(6,3)
\psline[linestyle=solid](-1,2)(6,2)
\psline[linestyle=solid](2,1)(6,1)
\psline[linestyle=solid](-1,4)(-1,1)
\psline[linestyle=solid](0,4)(0,1)
\psline[linestyle=solid](1,4)(1,1)
\psline[linestyle=solid](2,4)(2,1)
\psline[linestyle=solid](3,4)(3,0)
\psline[linestyle=solid](4,4)(4,0)
\psline[linestyle=solid](5,4)(5,0)
\psline[linestyle=solid](6,4)(6,0)
\psdot[dotstyle=x,dotsize=0.5](2.5,0.5)
\psline(-1,1)(2,1)(2,0)(6,0)
\pspolygon[fillstyle=hlines,linestyle=none](-1,1)(2,1)(2,0)(6,0)(6,-1)(-1,-1)
\rput(2.8,0.8){1} \rput(3.8,0.8){3}  \rput(4.8,0.8){7} \rput(5.8,0.8){13} 
\rput(0.8,1.8){8} \rput(1.8,1.8){4} \rput(2.8,1.8){2} \rput(3.8,1.8){6} \rput(4.8,1.8){12}  
\rput(1.8,2.8){9} \rput(2.8,2.8){5} \rput(3.8,2.8){11} 
\rput(2.8,3.8){10}
\end{pspicture}
\psset{unit=0.08\columnwidth}
\begin{pspicture}(-0.5,-0.5)(5.5,3.5)
\psline(2,3)(2,0) \psline(3,2)(3,0) \psline(4,1)(4,0) \psline(1,1)(1,0) 
\psline(2,3)(3,2)(4,1)(5,0)
\psline(2,2)(1,1)(0,0)
\psline(2,2)(3,1)(4,0)
\psline(1,0)(2,1)(3,0)
\psdots[dotstyle=o,dotsize=0.6](0,0)(2,0)(3,0)(4,0)(5,0)
\psdots[dotstyle=o,dotsize=0.6](2,1)(3,1)(4,1)
\psdots[dotstyle=o,dotsize=0.6](3,2)
\psdots[dotstyle=o,dotsize=0.6,fillstyle=solid,fillcolor=black!25](2,3)
\psdots[dotstyle=o,dotsize=0.6,fillstyle=solid,fillcolor=black!25](2,2)
\psdots[dotstyle=o,dotsize=0.6,fillstyle=solid,fillcolor=black!25](1,1)
\psdots[dotstyle=o,dotsize=0.6,fillstyle=solid,fillcolor=black!25](1,0)
\rput(2,3){1} 
\rput(2,2){2}\rput(3,2){3}
\rput(1,1){4}\rput(2,1){5}\rput(3,1){6}\rput(4,1){7}
\rput(0,0){8}\rput(1,0){9} 
\rput(2,0){10}\rput(3,0){11}\rput(4,0){12}\rput(5,0){13}
\end{pspicture}
\caption{Part of the square lattice that can be reached from animals
  up to size $4$ (left). The number of lattice animals of size $n
  \leq 4$ equals the number of subgraphs in the neighborhood graph (right)
  that contain vertex $1$. \label{fig:graph_redelmeier}}
\end{figure}
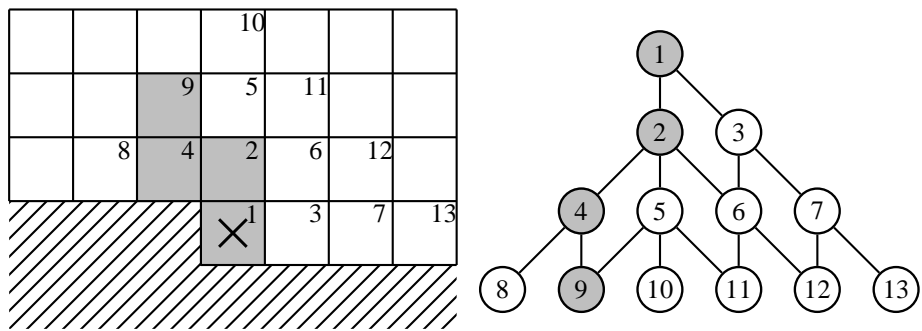

In order to break the translational symmetry we demand that the
initial site, which is contained in all animals, is an extremal site
with respect to the lexicographic order of lattice
coordinates. Figure~\ref{fig:graph_redelmeier} illustrates how this
can be achieved in the square lattice. We simply block all lattice
sites from further consideration that are in a row below the initial
site (marked with a cross) or in the same row and to the left of the
initial site. The generalization to $d>2$ is straightforward. These
blocked sites are never added to the untried set, but they need to be
taken into account when we compute the perimeter $t$. 


We start with all lattice sites being marked ``free'' or ``blocked,''
except for the initial site, which is marked ``counted.'' Furthermore
$n=1$, $t=1$ and the initial site being the only element of the
untried set $U$. Redelemeier's algorithm works by invoking the
following routine with this initial settings:
\begin{itemize}
\item Iterate until $U$ is empty:
\begin{enumerate}
\item \label{start} Remove a site $s$ from $U$.
\item $F := $ set of ``free'' neighbors, $B := $ set of
  ``blocked'' neighbors of $s$. $N := |U|+|B|$.
\item Count new cluster: increase $g_{n,t+N-1}$ by one.
\item If $n < \smax$:
  \begin{enumerate}
  \item Mark all sites in $F$ and $B$ as
    ``counted.''
 \item Call this routine recursively with 
  $U' = U\cup F$, $n' = n+1$ and $t' = t+N-1$.
\item Relabel sites in $F$ as ``free'' and sites in $B$ as
  ``blocked.''
  \end{enumerate}
\end{enumerate}
\item{Return.}
\end{itemize}

Since the algorithm generates each lattice animal explicitely, its
running time scales like $A_d(n)$. This exponential complexity implies
hard limits for the accessible animal sizes.  All we can do is to keep
the prefactor in the time complexity function small, i.e., to implement
each step of Redelemeier's algorithm as efficiently as possible by
using an appropriate data structure for the untried set and for the
lattice, see \cite{mertens:90}. 

Another crucial element of tuning Redelmeier's algorithm is the computation of
the neighborhood of a lattice cell. In a recent paper
\cite{aleksandrowicz:barequet:09a}, Aleksandrowicz and Barequet
observed that Redelmeier's algorithm can be interpreted as the
counting of subgraphs in a graph that represents the neighborhood
relation of the lattice. Figure~\ref{fig:graph_redelmeier} illustrates
this for the square lattice.

For any lattice, the neighborhood graph can be precomputed and be
represented as an adjacency list which is then fed to the actual
subgraph counting algorithm. That way the computation of the
neighbors of a lattice cell is taken out of the counting loop, and the
prefactor  in the exponential scaling is reduced.

The size of the neighborhood graph or, equivalently, the number of lattice points
required to host lattice animals of size $n$ scales like
$\Theta(n^d)$. Aleksandrowicz and Barequet
\cite{aleksandrowicz:barequet:09a} claimed that this exponential growth of memory
with $d$ represents a serious bottleneck for Redelmeier's algorithm in
high dimensions. In a subsequent paper
\cite{aleksandrowicz:barequet:09b}, they therefore present a variation of
the algorithm that avoids the storage of the full graph by 
computing the relevant parts of the graph on demand. This cuts down
the space complexity to a low order polynomial in $d$, but it
forfeits the gain in speed that can be obtained by precomputing the
complete neighborhood graph.

We claim that in practice the space complexity of Redelmeier's
algorithm is no bottleneck. The reason is that the
prefactor in the $\Theta(n^d)$ scaling can be made small enough to
hold the complete graph in memory for all values of $n$ and $d$ for
which $A_d(n)$ is computable in reasonable time. 

\begin{figure}
\psset{xunit=0.06\columnwidth,yunit=0.04\columnwidth}
\begin{center}
\begin{pspicture}(-3,-0.6)(12.5,12.5)
  \psaxes[ylogBase=10,Dy=2]{->}(0,0)(0,0)(12.5,12.5)
  \uput[0](12.5,0){$n$}
  \psset{showpoints=true,plotstyle=curve,dotsize=6pt}
  \dataplot[dotstyle=o]{\crystalballdata}
  \dataplot[dotstyle=Square]{\crystalcubedata}
\end{pspicture}
\end{center}
\caption{\label{fig:volumes}Volume needed to cage animals of size
  $n$ on the hypercubic lattice of dimension $d=9$. Cubic
  ($\square$) versus spherical cages ({\tiny$\bigcirc$}). }
\end{figure}
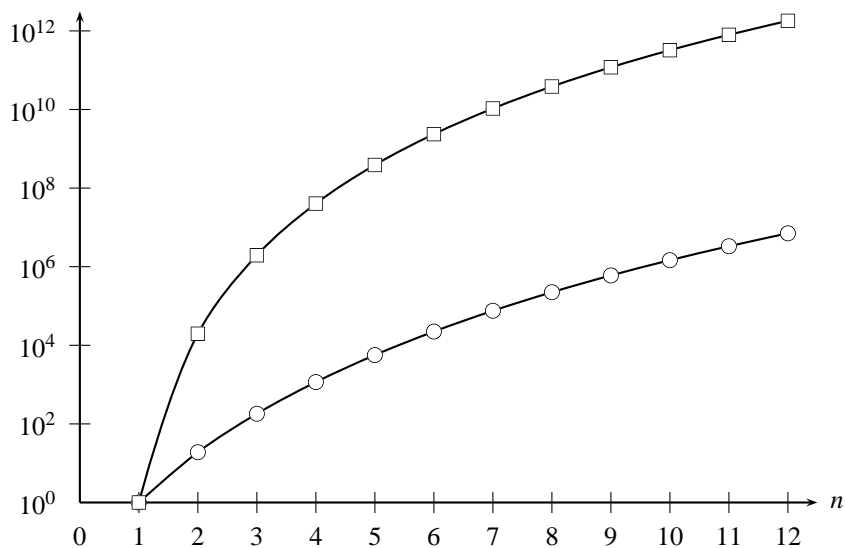

The key observation is that in \cite{aleksandrowicz:barequet:09a},
Aleksandrowicz and Barequet used a hyper-cube of side length $2n$ of the lattice to
host the lattice animals, whereas a hyper-sphere of radius $n$
suffices. This is a significant difference, as can be seen from the
analogous situation 
in $\mathbb{R}^d$. Here the volume of a cube is much larger
than the volume of the inscribed Euclidean sphere,
\begin{equation}
  \label{eq:cont-volume}
  \frac{\mbox{volume hypercube}}{\mbox{volume inscribed hypersphere}}
  = \frac{2^d\Gamma(\frac{d}{2}+1)}{\pi^{d/2}}
  \simeq \left(\frac{2d}{\pi\mathrm{e}}\right)^{d/2}\frac{1}{\sqrt{\pi d}}\,.
\end{equation}
In the lattice $\mathbb{Z}^d$, the number of lattice points in a cube 
of side length $2n+1$ is 
\begin{equation}
  \label{eq:cube-volume}
  C(d,n) = (2n+1)^d\,.
\end{equation}
Let $B(d,n)$ denote the number of lattice sites that are $n$ steps or
less away from the origin. This ``volume of the crystal ball'' can be computed
recursively via 
\begin{eqnarray}
  \label{eq:ball-volume}
  B(1,n) &=& 2n+1 \nonumber\\
  B(d,n) &=& B(d-1,n) + 2\sum_{k=0}^{n-1} B(d-1,k)\,,
\end{eqnarray}
which reflects the fact that the crystall ball in dimension $d$ can
be decomposed into $(d-1)$-dimensional slices whose diameter decreases
with increasing distance from the central slice. The recursion
\eqref{eq:ball-volume} tells us that $B(d,n)$ is a polynomial in $n$
of degree $d$ which can easily be computed, see \seqnum{A001845} to
\seqnum{A001848} on oeis.org for the polynomials for $d=3,\ldots,6$.
Note that $B(d,n)$ can also be computed through the 
generating function \cite{conway:sloane:97}
\begin{equation}
  \label{eq:ball-generating}
  \frac{(1+x)^d}{(1-x)^{d+1}} = \sum_{n=0}^\infty B(d,n) x^n\,.
\end{equation}

The number of lattice sites needed to cage lattice animals of size $n$ is
very close to $B(d,n-1)/2$ for spherical cages and $C(d,n-1)/2$ for
cubical cages. Figure~\ref{fig:volumes} shows both numbers for the case $d=9$.
As you can see, $d=9$ and $n=12$ requires memory on the Terabyte scale
if one uses cubic cages, but only a few Megabytes for spherical cages.

\section{Performance}
\label{sec:performance}

\begin{table}
\centering
\begin{tabular}{|r|c|c||r|c|}
\hline
& \multicolumn{2}{c||}{Perimeter Polynomial} & \multicolumn{2}{c|}{$A_d(n)$}\\\hline
$d$ & old $\smax$ & new $\smax$ & old
$\smax$ &  new $\smax$\\
\hline
3 & 15 & 18 & 18 & 19\\
4 & 10 & 15 & 15 & 16\\
5 & \phantom{1}9 & 14 & 13 & 15\\
6 & \phantom{1}8 & 14 & 10 & 15\\
7 & \phantom{1}8 & 13 & 10 & 14\\
8 &  &  11 & 8 & 12\\
9 &  &  11 & 4 & 12\\
10 & & 11 & & 12\\
\hline
\end{tabular}
\caption{Range of perimeter polynomials and animal numbers in dimensions
  $d\geq 3$ that have been found by exhaustive enumerations. The old
  perimeter polynomials are from \cite{mertens:90} and
  \cite{gaunt:sykes:ruskin:76}, the old values of $A_d(n)$ are from
  \cite{aleksandrowicz:barequet:09a} and \cite{aleksandrowicz:barequet:09b}. 
\label{tab:new_limits}}
\end{table}

Our implementation of the Redelmeier algorithm consists of two programs.
The first program computes the neighborhood graph
of a specified lattice and writes this graph as an adjacency list into
a file. The second program reads this
file and computes the corresponding perimeter polynomials.
The programs are written in C++ and 
can be downloaded from the project webpage  \cite{animals:site}.

When run on a laptop with an Intel\textregistered\ Core\texttrademark\
2 Duo CPU at 2 GHz, the program enumerates perimeter polynomials at a
rate of roughly $2 \cdot 10^7$ animals per second. This means that
generating and counting one lattice animal and measuring its perimeter
takes about $100$ clock cycles, which is reasonable for a program
compiled from C++.

At this rate, our laptop needs 35 days to enumerate the perimeter
polynomials for $d=9$ and $n\leq 11$ (see Table~\ref{tab:numbers}).
Computing the next perimeter polynomial ($n=12$) would take more
than three years. Note that according to Figure~\ref{fig:volumes}, the
neighborhood graph for $d=9$ and $n=12$ easily fits into the
memory of a run-of-the-mill laptop. These numbers illustrate that for
for all practical purposes, the bottleneck of Redelemeier's algorithm
is time, not memory. 

\begin{table}
\begin{tabular}{|c|r|r|r|r|}
\hline
$n$ & $A_6(n)$ & $A_7(n)$ & $A_8(n)$ & $A_9(n)$ \\
\hline
1 & 1 & 1 & 1 & 1 \\
2 & 6 & 7 & 8 & 9 \\
3 & 66 & 91 & 120 & 153 \\
4 & 901 & 1484 & 2276 & 3309 \\
5 & 13881 & 27468 & 49204 & {\bf 81837} \\
6 & 231008 & 551313 & 1156688 & {\bf 2205489} \\
7 & 4057660 & 11710328 & 28831384 & {\bf 63113061} \\
8 & 74174927 & 259379101 & 750455268 & {\bf 1887993993} \\
9 & 1398295989 & 5933702467 &  {\bf 20196669078} & {\bf 58441956579} \\
10 & 27012396022 &139272913892 &  {\bf 558157620384} & {\bf 1858846428437} \\
11 & {\bf 532327974882}  & {\bf 3338026689018} &  {\bf 15762232227968} & {\bf 60445700665383} \\
12 & {\bf 10665521789203}  & {\bf 81406063278113} &  {\bf
  453181069339660} &  {\bf 2001985304489169}\\
13 & {\bf 227093585071305} & {\bf 2014611366114053} & & \\
14 & {\bf 4455636282185802} & {\bf 50486299825273271} & & \\
15 & {\bf 92567760074841818} & & & \\\hline
\end{tabular}
\caption{Number of lattice animals in the hypercubic lattice for $d =
  6\ldots9$ obtained by \emph{direct enumeration}. New results in boldface, the numbers of smaller animals
  are from
  \cite{aleksandrowicz:barequet:09a,aleksandrowicz:barequet:09b} and
  references therein.  Note that in \cite{gaunt:80}, $A_8(n)$ and $A_9(n)$ for $n\leq 9$ were
  computed rather than enumerated by the same method that we will use
  in Section~\ref{sec:combinatorics} to extend this table to $n\leq
  14$ and all values of $d$.}\label{tab:numbers}
\end{table}

Using a parallel implementation \cite{mertens:lautenbacher:92} that we
ran on a Linux cluster with 128 Intel\textregistered\
Xeon\textregistered\ 3.2 GHz CPUs, or for the most demanding
computations, on a SciCortex SC5832 with 972 MIPS64 6-core nodes, we
could extend the table of known perimeter polynomials and animal
counts considerably, see Table~\ref{tab:new_limits}. The new values
for $d \leq 5$ are
\begin{eqnarray*}
  A_3(19) &=& {\bf \phantom{3\,}651\,459\,315\,795\,897}\,,\\
  A_4(16) &=& {\bf \phantom{3\,}692\,095\,652\,493\,483}\,,\\
  A_5(14) &=& {\bf \phantom{3\,}227\,093\,585\,071\,305}\,,\\
  A_5(15) &=& {\bf 3\,689\,707\,621\,144\,614}\,.
\end{eqnarray*}
The numbers for $6\leq d \leq 9$ are given in Table~\ref{tab:numbers},
the corresponding perimeter polynomials
can be found on the project webpage \cite{animals:site}. Before we
evaluate the results, we will discuss a combinatorial argument
that allows us to extend the enumeration data considerably.

Note that the most demanding computation in this paper
was the enumeration of the perimeter polynomial for $n = 14$ in
$d=6$. On a single core of a MIPS64, this enumeration would have taken
77 CPU years, on our Laptop from above it would still have taken about
7 CPU years. In practice we used a parallel implementation that ran on
many cores (and several different machines) such that no computation
took longer than two weeks wall clock time.

\section{Proper Animals}
\label{sec:combinatorics}

A lattice animal of size $n$ can't span more than $n-1$
dimensions. This simple observation allows us to derive explicit
formulas for $A_d(n)$ for fixed $n$. Obviously $A_d(1) = 1$ and
$A_d(2)=d$. A lattice animal of size $n=3$ is either a one-dimensional
``stick'' with $d$ possible orientations or ``L-shaped" and spanning
$2$ out of $d$ dimensions. Within these $2$ dimensions there are $4$
possible orientations for the L-shaped animal (see Figure~\ref{fig:polycubes}), hence
\begin{displaymath}
  A_d(3) = d + 4{d \choose 2} = 2d^2-d\,.
\end{displaymath}
For $n=4$, we have again the ``stick'' that lives in one dimension,
$17$ animals that span $2$ dimensions and $32$ animals that span $3$
dimensions:
\begin{displaymath}
  A_d(4) = d + 17{d \choose 2} + 32{d \choose 3} = \frac{16}{3}d^3-\frac{15}{2}d^2+\frac{19}{6}d\,.
\end{displaymath}
In general we can write
\begin{equation}
  \label{eq:lunnon}
  A_d(n) = \sum_{i=0}^d {d \choose i} \DX(n,i)\,,
\end{equation}
where $\DX(n,i)$ denotes the number of fixed \emph{proper} animals of
size $n$ in dimension $i$. An animal is called proper in dimension $d$
if it spans all $d$ dimensions. Equation~\eqref{eq:lunnon}  is due to Lunnon
\cite{lunnon:75}. If we know $A_d(n)$ for a given $n$ and $d
\leq \dmax$, we can use \eqref{eq:lunnon} to compute $\DX(n,d)$ for
the same value of $n$ and all $d\leq \dmax$, and vice versa. 

Since $\DX(n,i) = 0$ for $i\geq n$, Lunnon's equation tells us that
$A_d(n)$ is a polynomial of degree $n-1$ in $d$, and since $A_0(n) =
0$ for $n>1$, it suffices to know the values $A_1(n), A_2(n), \ldots,
A_{n-1}(n)$ to compute the polynomial $A_d(n)$. From our enumeration
data (Table~\ref{tab:new_limits}), we can compute these polynomials up
to $A_d(11)$, see Table~\ref{tab:A-polynomials}.

\begin{table}
{\small
\begin{eqnarray*}
  A_d(2) &=& d\\
  A_d(3) &=& 2d^2-d\\
  A_d(4) &=&\frac{16}{3}d^3-\frac{15}{2}d^2+\frac{19}{6}d\\
 A_d(5) &=&\frac{50}{3}d^4-42 d^3+\frac{239}{6}d^2-\frac{27}{2}d\\
 A_d(6) &=&\frac{288}{5}d^5-216 d^4+\frac{986}{3}d^3-231
 d^2+\frac{926}{15}d\\
A_d(7) &=&\frac{9604}{45}d^6-1078
d^5+\frac{20651}{9}d^4-\frac{14927}{6}d^3+\frac{120107}{90}d^2-\frac{827}{3}d\\
A_d(8) &=&
\frac{262144}{315}d^7-\frac{26624}{5}d^6+\frac{132320}{9}d^5-\frac{65491}{3}d^4+\frac{1615991}{90}d^3
-\frac{113788}{15}d^2+\frac{52589}{42}d\\
A_d(9) &=& \frac{118098}{35}d^8-26244
d^7+\frac{447903}{5}d^6-\frac{511082}{3}d^5+\frac{23014949}{120}d^4-\frac{1522261}{12}
d^3\\
& & +\frac{38839021}{840}d^2-\frac{30089}{4}d\\
A_d(10) &=& \frac{8000000}{567}d^9-\frac{2720000}{21} d^8
+\frac{14272000}{27}d^7-\frac{11092360}{9} d^6+\frac{239850598}{135}d^5\\
& & -\frac{14606026}{9}d^4+\frac{1067389643}{1134} d^3-\frac{42595493}{126}d^2+\frac{2804704}{45}d\\
  A_d(11) &=& \frac{857435524}{14175}d^{10}-\frac{67319318}{105} d^9
  +\frac{2884481974}{945} d^8 -\frac{380707987}{45} d^7 
   +\frac{40341440233}{2700}d^6\\ 
   & &-\frac{1260803635}{72}d^5+\frac{79118446751}{5670}d^4
   -\frac{19252021283}{2520}d^3+\frac{17126616179}{6300}d^2-\frac{7115086}{15}d\\
A_d(12) &=&
\frac{509607936}{1925}d^{11}-\frac{15925248}{5}d^{10}+\frac{607592448}{35}d^9-\frac{1956324864}{35}d^8+\frac{2930444704}{25}d^7\\
& &
-\frac{2522387284}{15}d^6+\frac{17894522696}{105}d^5-\frac{1242881121}{10}d^4\\
& & +\frac{22272055467}{350}d^3-\frac{4225468993}{210}d^2+\frac{181356011}{66}d\\
A_d(13) &=&\frac{551433967396}{467775}d^{12}-\frac{75047226332}{4725}d^{11}+\frac{166095324499}{1701}d^{10}-\frac{48436628461}{135}d^9\\
 & &+\frac{49499551181119}{56700}d^8-\frac{1335959158369}{900}d^7+\frac{248648897740349}{136080}d^6-\frac{25156285613453}{15120}d^5\\
&
&+\frac{757565736903221}{680400}d^4-\frac{5607318230581}{10800}d^3+\frac{12648671104037}{83160}d^2-\frac{135165335}{6}d\\
A_d(14) &=& \frac{4628074479616}{868725}d^{13}
-\frac{23612624896}{297}d^{12}
+\frac{3309261190144}{6075}d^{11}
-\frac{304034058496}{135}d^{10}\\
& &
+\frac{12648090831712}{2025}d^9
-\frac{553376997376}{45}d^8
+\frac{758347226205724}{42525}d^7
-\frac{2633038200122}{135}d^6\\
& &+\frac{98388569956577}{6075}d^5
-\frac{2734657007119}{270}d^4
+\frac{11824147558382}{2475}d^3
-\frac{560344373791}{330}d^2\\
& &+\frac{97500388612}{273}d
\end{eqnarray*}}
\caption{Number of lattice animals of sizes $2,\ldots,14$ in hypercubic
  lattices of dimension $d$. The polynomials for $n\leq 11$ have been
  obtained by direct enumeration and confirm those
  listed in \cite{barequet:barequet:rote:10}. Polynomials for $n>11$
  have been computed from enumeration data and known values of
  $\DX(n,n-k)$. \label{tab:A-polynomials}}
\end{table}

In order to compute $A_d(12)$, we need to know $A_{11}(12)$ or
equivalently, $\DX(12,11)$. The latter can actually be computed
with pencil and paper. That's because an animal of size $12$ in $11$
dimensions has to span a new dimension with each of its cells to be
proper. In particular, its cells can't form loops. Hence computing
$\DX(12,11)$ is an exercise in counting trees. This is
true for $\DX(n,n-1)$ in general, so let's compute this function.

The adjacency graph of a lattice animal of size
$n$ is an edge labeled graph with $n$ vertices, in which each vertex represents a
cell of the animal and two vertices are connected if the corresponding
cells are neighbors in the animal. Every edge of the
adjacency graph is labeled with the dimension along which the two
cells touch each other.

In the case $\DX(n,n-1)$, every pair of adjacent cells must span a new
dimension. Therefore the corresponding adjacency graph contains
exactly $n-1$ edges, i.e. it is a tree, and each edge has a unique
label. There are two directions for each dimensions that we represent
by the orientation of the edge in the tree. Hence $\DX(n,n-1)$ equals
the number of directed, edge-labeled trees of size $n$, where in our
context ``directed'' means that each edge has an arbitrary orientation
in addition to its label. 

The number of \emph{vertex} labeled trees of size $n$ is given
by $n^{n-2}$, the famous fomula published by Cayley in 1889
\cite{cayley:1889}.  The number of \emph{edge} labeled trees
seems to be much less known, at least it is proven afresh in recent papers
like \cite{cameron:95}.  The following nice derivation is from
\cite{barequet:barequet:rote:10}. Start with a
vertex labeled tree of size $n$ and mark the vertex with label $n$ as
the root. Then shift every label smaller than $n$ from its vertex to the incident
edge towards the root. This gives an edge labeled tree with a single
vertex marked (the root). Since the mark can be on any vertex, the number
of edge labeled trees equals the number of vertex labeled trees
divided by the number of vertices. According to Cayley's formula, this
number is $n^{n-3}$. And since each directed edge can have two
directions, we get
\begin{equation}
  \label{eq:dx1}
  \DX(n,n-1) = 2^{n-1}\,n^{n-3}\,.
\end{equation}
This formula has been known in
the statistical physics community for a long time
\cite{fisher:essam:61}.  We used it to compute $\DX(12,11)$ and then
$A_d(12)$ (Table~\ref{tab:A-polynomials}).

We can proceed further along this line. To compute $A_d(13)$ we have to
extend our enumeration data by $A_{12}(13),\ldots,A_{8}(13)$ or
equivalently by $\DX(13,12),\ldots,\DX(13,8)$. 
What we need are formulas $\DX(n,n-k)$ for $k>1$.

For $k>1$, there is no longer a simple correspondence between
edge labeled trees and proper animals. 
We need to take into account that
there are edge labels with the same value,
that the adjacency graph may contain loops, and that some labeled
trees represent a self-overlapping
and therefore illegal lattice animal. A careful consideration of
these issues yields
\begin{equation}
  \label{eq:dx2}
  \DX(n,n-2) = 2^{n-3}\,n^{n-5} (n-2) (9-6n+2 n^2)\,,
\end{equation}
see \cite{barequet:barequet:rote:10} for the derivation of \eqref{eq:dx2}.

For $k>2$, the computation of $\DX(n,n-k)$ gets very complicated and is better left to a
computer.  In \ref{sec:dx-structure} we show that 
\begin{equation}
  \label{eq:dx-form}
  \DX(n,n-k) = 2^{n-2k+1} n^{n-2k-1}\,g_k(n)\,,
\end{equation}
where $g_k(n)$ is a polynomial of degree $3k-3$.  Hence we can compute
$g_k$ from $3k-2$ data points, like the values of $\DX(n,n-k)$ for
$n=k,\ldots,4k-3$.  Our enumeration data suffices to compute $g_2$ and
$g_3$ with this method, but not $g_4$.

However, there is a trick that allows us to compute
$g_k$ from many fewer data points. The free energy 
\begin{displaymath}
  f_n = \frac{1}{n}\log A_d(n)
\end{displaymath} 
has a well defined $1/d$ expansion whose coefficients depend on
$n$. If we \emph{assume} that these coefficients are bounded in the
limit $n\to\infty$, most of the coefficients in $g_k$ are fixed, and we only need to know
$k+1$ data points to fully determine $g_k$.  See \ref{sec:dx-coeff} 
for the details of this argument. In our case this enables us to compute $g_k$
up to $k=7$, see Table~\ref{tab:gk},  and consequently
$A_d(13)$ and $A_d(14)$, see Table~\ref{tab:A-polynomials}.

\begin{table}
  {\small
  \begin{eqnarray*}
    g_2(n) &=& (n-2)(9-6n+2n^2)\\
    g_3(n) &=& \frac{n-3}{6}(-1560 + 1122n - 679n^2 + 360n^3 - 104n^4
    + 12n^5) \\
    g_4(n) &=& \frac{n-4}{6}(204960 - 114302n + 41527n^2 - 17523n^3 + 7404n^4 - 2930n^5 + 
 828n^6 - 128n^7 + 8n^8)\\
    g_5(n) &=& \frac{n-5}{360}(-3731495040 + 1923269040n - 535510740n^2 + 150403080n^3 - 
  42322743n^4\\
    & & \phantom{n-5(} + 12397445n^5  -4062240n^6 + 1335320n^7 - 
  356232n^8 + 62240n^9 - 6000n^{10} + 240n^{11}) \\
   g_6(n) &=& \frac{n-6}{360}(1785362705280 - 939451308048n + 248868418932n^2 - 56265094748n^3 \\
   & & \phantom{n-6(} +11984445891n^4-2448081038n^5 + 535284255n^6 - 127651774n^7 + 
  33940138n^8\\
   & &  \phantom{n-6(}  - 9580440n^9 + 2398912n^{10}-440688n^{11} + 
  51856n^{12} - 3424n^{13} + 96n^{14})\\
  g_7(n) &=&
  \frac{n-7}{45360}(-156017752081551360+85163968967728896n-22517704978919136n^2\\
   & &\phantom{45360(}+4585470174542376n^3-851686123590540n^4+146137469433102n^5\\
   & &\phantom{45360(}-24441080660523n^6+4148836864606n^7-747463726205n^8\\
   & &\phantom{45360(}+149724735468n^9-33793043592n^{10}+8322494124n^{11}-1946680944n^{12}\\
   & &\phantom{45360(}+363148352n^{13}-47679184n^{14}+4019904n^{15}-193536n^{16}+4032n^{17})
 \end{eqnarray*}}
  \vspace{-5ex}
  \caption{Polynomials $g_k(n)$ that appear in $\DX(n,n-k)$
    \eqref{eq:dx-form}.  The polynomials $g_2,\ldots,g_6$ can be found
    as $g_{k,0}$ in Appendix 2 of \cite{peard:gaunt:95}. As far as we
    know, the polynomial $g_7$ has not been published before. See the
    Appendix of this paper for the method how to compute the $g_k$.}
  \label{tab:gk} 
\end{table}

We actually know all data to compute $A_d(15)$ with the exception of
the number $A_7(15)$. On our laptop, the enumeration of the missing number
$A_7(15)$ would take about 80 years.  On a parallel system with a few
hundred CPUs this would still take several months, which is not out of
reach. Computing the formula for $A_d(16)$, however, is
definitely beyond the power of our machinery.

Before we turn our attention to the analysis of the enumeration data
we note that Lunnon's equation \eqref{eq:lunnon} has a corresponding equation
for perimeter polynomials:
\begin{equation}
  \label{eq:lunnon-g}
  g_{n,t}^{(d)} = \sum_{i} {d \choose i} G_{n,t-2(d-i)n}^{(i)}\,.
\end{equation} 
$G_{n,t}^{(d)}$ denotes the number of proper $d$-dimensional animals of size $n$
and perimeter $t$. Since $G_{n,t}^{(d)} = 0$ for $d > n-1$,
we can write
\begin{equation}
  \label{eq:Pdnq}
  P_d(n,q) = q^{2dn-2(n-1)} \sum_{i=1}^{n-1} {d \choose i} \sum_t
  G_{n,t}^{(i)} q^{t-2-2n(i-1)}\,.
\end{equation}
For a given value of $n$, \eqref{eq:Pdnq} represents the perimeter polynomial
for general dimension $d$. Our enumeration data allowed us
to compute the $G_{n,t}^{(d)}$ and hence the formulas \eqref{eq:Pdnq}
for $n\leq 11$ (see \cite{animals:site} for the data), extending the 
previously known formulas for $n\leq 7$ \cite{gaunt:sykes:ruskin:76}.
A computation of the next formula $P_d(12,q)$ requires the knowledge
of the perimeter polynomials for $d\leq11$ and $n\leq12$. The
enumeration for $d=11$ and $n=12$ alone would take
ca. 38 years on our laptop.

\section{Mean Cluster Size}

From the perimeter polynomials we can compute moments of the
cluster statistics like the mean cluster size
\begin{equation}
  \label{eq:def-S}
  S(p) = \frac{1}{p} \sum_{n=1}^\infty n^2 p^n P_d(n,1-p) =
  \sum_{r} b_d(r) p^r\,. 
\end{equation}
The coefficients of the series expansion read
\begin{eqnarray}
  \label{eq:S-series-1}
  b_d(r) &=& \sum_{n=1}^{r+1} n^2 \sum_t g_{n,t}^{(d)} {t \choose
    {r+1-n}} (-1)^{r+1-n} \nonumber\\
  &=& (r+1)^2 A_d(r+1) + \sum_{n=1}^{r} n^2 \sum_t g_{n,t}^{(d)} {t \choose
    {r+1-n}} (-1)^{r+1-n}\,. 
\end{eqnarray}
Since we can compute $A_d(r+1)$ from the perimeter polynomials
$P_d(n,q)$ for $n\leq r$ via \eqref{eq:extra-A-from-P},  we can also
compute the series coefficients $b_d(r)$ from this set of
perimeter polynomials. If we happen to know $A_d(r+2)$, we can get an extra
coefficient through
\begin{equation}
  \label{eq:S-series-2}
  \eqalign{b_d(r+1) = &(r+2) A_d(r+2)  \\
  &+\sum_{n=1}^r n(n-r-1) \sum_t
  g_{n,t}^{(d)} {t \choose
    {r+2-n}} (-1)^{r-n}\,.}
\end{equation}
This formula can be derived by solving \eqref{eq:extra-A-from-P} for
$\sum_t g_{n,t} t$ and plugging the result into
\eqref{eq:S-series-1}. 

\begin{table}
{
\begin{tabular}{|c|r|r|r|r|r|}
\hline
r & $d = 3$ & $d = 4$ & $d = 5$ & $d = 6$\\
\hline
1 & 6 & 8 & 10 & 12\\
2 & 30 & 56 & 90 & 132\\
3 & 114 & 320 & 690 & 1272\\
4 & 438 & 1832 & 5290 & 12252\\
5 & 1542 & 9944 & 39210 & 115332\\
6 & 5754 & 55184 & 293570 & 1091472\\
7 & 19574 & 290104 & 2135370 & 10159252\\
8 & 71958 & 1596952 & 15839690 & 95435172\\
9 & 233574 & 8237616 & 113998170 & {\bf 883192392}\\
10 & 870666 & 45100208 & {\bf 840643170} & {\bf 8258076192}\\
11 & 2696274 & {\bf 229502616} & {\bf 6017266290} & {\bf 76196541732}\\
12 & 10375770 & {\bf 1254330128} & {\bf 44178511010} & {\bf
  710151162432}\\
13 & 30198116 & {\bf 6307973352} & {\bf 315024296150} & {\bf
  6540805549192}\\
14 & {\bf 122634404} & {\bf 34574952952} & {\bf 2307462163110} &  {\bf
  60831844077672}\\
15 & {\bf 327024444} & {\bf 171364602736} &  & \\
16 & {\bf 1460721616} &  &  & \\
17 & {\bf 3347244554} &  &  & \\
18 & {\bf 17795165832} & & &\\
\hline
\end{tabular}

\vspace{2ex}
\begin{tabular}{|c|r|r|r|}
\hline
r & $d=7$ & $d=8$ & $d = 9$\\
\hline
1 & 14 & {\bf 16} & {\bf 18} \\
2 & 182 & {\bf 240} & {\bf 306} \\
3 & 2114 & {\bf 3264} & {\bf 4770} \\
4 & 24542 & {\bf 44368} & {\bf 74322} \\
5 & 280238 & {\bf 595632} & {\bf 1146834} \\
6 & 3210074 & {\bf 8012384} & {\bf 17720514} \\
7 & 36394302 & {\bf 107053424} & {\bf 272530194} \\
8 & 414610014 & {\bf 1434259248} & {\bf 4198328082} \\
9 & {\bf 4685293438} & {\bf 19125485024} & {\bf 64487361906} \\
10 & {\bf 53201681162} & {\bf 255662267296} & {\bf 991886672898} \\
11 & {\bf 600207546946} & {\bf 3405928921264} & {\bf 1522687319670} \\
12 & {\bf 6800785109594} & {\bf 45466350310880} & {\bf 233996383280898}\\
13 & {\bf 76649757121000} & & \\
\hline
\end{tabular}
}
\caption{Series coefficients of the mean cluster size  $S(p) =
  \sum\limits_r{b_r p^r}$ in hypercubic lattices of dimension
  $d$. New values in boldface, older values from \cite{mertens:90}
  ($d=3$) and \cite{gaunt:sykes:ruskin:76} ($d=4,\ldots,7$)
  and references therein. \label{tab:series-S}}
\end{table}

We used \eqref{eq:S-series-2} to extend the table of coefficients (Table~\ref{tab:series-S}) for
$d\geq 8$, since here we know perimeter polynomials only up to $n=11$,
but the cluster numbers $A_d$ up to $n=14$.

 

\section{Growth Rates and Exponents}
\label{sec:lambda}

\begin{figure}
\psset{xunit=8\columnwidth,yunit=11\columnwidth}
\begin{center}
\begin{pspicture}(-0.03,3.663)(0.085,3.703)
  \psaxes[Oy=3.67,Dx=0.02,Dy=0.01]{->}(0.00,3.67)(0.08,3.703)
  \uput{25pt}[270](0.04,3.67){$1/n^{\Delta+1}$}
  \uput{35pt}[180]{90}(0,3.685){$\log\lambda_9(n)$}
  \psplot[linestyle=solid]{0.0}{0.075}{3.70057 0.400082 x mul sub}
  \psset{showpoints=true,plotstyle=dots,dotsize=6pt}
  \dataplot[dotstyle=o]{\lambdaninedata}
\end{pspicture}
\end{center}
\caption{\label{fig:lambda}Growth rate $\lambda_d$ for $d=9$. The
  symbols are $\lambda_9(n)$ computed from
  Eqs.~\eqref{eq:lambda-theta-system}. The correction exponent
  $\Delta=0.58$ and the line are the result of a numerical fit to the
  three leftmost data points.}
\end{figure}
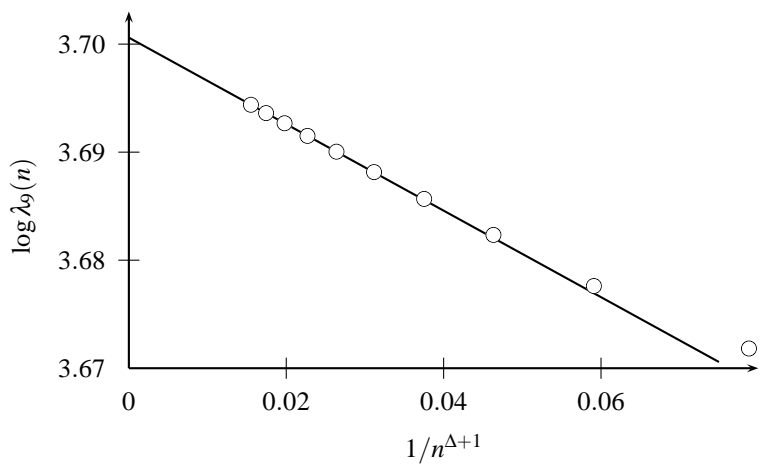

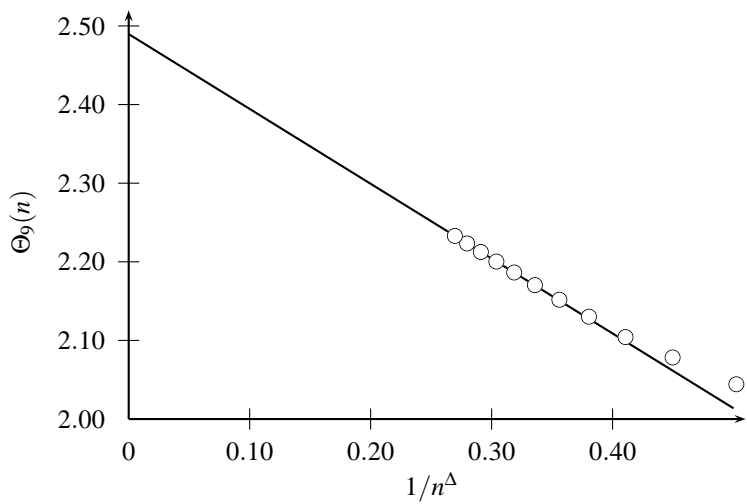
\begin{figure}
\psset{xunit=1.23\columnwidth,yunit=0.8\columnwidth}
\begin{center}
\begin{pspicture}(-0.16,1.925)(0.5,2.52)
  \psaxes[Oy=2.00,Dx=0.10,Dy=0.10]{->}(0.0,2.00)(0.51,2.52)
  \uput{20pt}[270](0.25,2.00){$1/n^\Delta$}
  \uput{35pt}[180]{90}(0,2.25){$\Theta_9(n)$}
  \psplot[linestyle=solid]{0.0}{0.5}{2.48948 0.951994 x mul sub}
  \psset{showpoints=true,plotstyle=dots,dotsize=6pt}
  \dataplot[dotstyle=o]{\thetaninedata}
\end{pspicture}
\end{center}
\caption{\label{fig:theta}Exponent $\Theta_d$ for $d=9$. The
  symbols are $\theta_9(n)$ computed from
  Eqs.~\eqref{eq:lambda-theta-system}. The correction exponent
  $\Delta=0.50$ and the line are the result of a numerical fit to the
  three leftmost data points. }
\end{figure}

The cluster numbers $A_d(n)$ 
are expected to grow asymptocically as
\begin{equation}
  \label{eq:scaling}
  A_d(n) \sim C \lambda_d^n\, n^{-\Theta_d}\,(1+\frac{b}{n^{\Delta}}
  + \mbox{corrections}),
\end{equation}
where the exponents $\Theta_d$ and $\Delta$ are universal constants,
i.e., their value depends on the dimension $d$, but not on the
underlying lattice, while $C$ and $b$ are nonuniversal, lattice
dependent quantities \cite{adler:etal:88}.  The universality
facilitates the computation of $\Theta_d$ for some values of $d$ using
field theoretic arguments. We know $\Theta_3=3/2$
\cite{parisi:sourlas:81,imbrie:03}, $\Theta_4=11/6$ \cite{dhar:83} and
$\Theta_d=5/2$ (the value for the Bethe lattice) for $d\geq\dc=8$, the
critical dimension for animal growth \cite{lubensky:isaacson:79}.  

The enumeration data for $A_d(n)$ can be used to estimate both $\lambda_d$
and $\Theta_d$.  For that we compute $\lambda_d(n)$ and $\Theta_d(n)$ as the
solutions of the system
\begin{equation}
  \label{eq:lambda-theta-system}
  \log A_d(n-k) = \log C + (n-k)\log\lambda_d(n) - \Theta_d(n) \log(n-k)
\end{equation}
for $k=0,1,2$.  We need three equations to eliminate the constant
$\log C$. Growth rate $\lambda_d$ and exponent $\Theta_d$ are
obtained by extrapolating the numbers $\lambda_d(n)$ and
$\Theta_d(n)$ to $n\to\infty$. From \eqref{eq:scaling} we expect
that
\begin{equation}
  \label{eq:lambda-finite-size}
  \log\lambda_d(n) \sim \log\lambda_d + \frac{b}{n^{\Delta+1}} 
\end{equation}
for large values of $n$. We used the data points $\lambda_d(n)$
for the three largest values of $n$ to fit the parameters $\log\lambda_d$,
$b$ and $\Delta$ in \eqref{eq:lambda-finite-size}. A plot of $\log\lambda_d(n)$
versus $n^{-\Delta-1}$ (Figure~\ref{fig:lambda}) then shows that the
data points in fact scale like \eqref{eq:lambda-finite-size}. The
resulting estimates for $\log\lambda_d$ are listed in
Table~\ref{tab:lambda}. They agree very well with the high precision
values from large
scale Monte Carlo simulations
\cite{hsu:nadler:grassberger:05,hsu:nadler:grassberger:05a}.

\begin{table}
  \centering
  \begin{tabular}{c|cl|cc}
    & \multicolumn{2}{c|}{\rule[-1.5ex]{0pt}{3ex}$\log \lambda_d$} &
    \multicolumn{2}{c}{$\Theta_d$} \\
    $d$ & enum. & \multicolumn{1}{c|}{MC} & enum. & exact, MC \\\hline
      3   & \rule{0pt}{2.5ex}2.12169  & 2.1218588(25)
      & 1.489 & $3/2$ \\
      4   & 2.58750 & 2.587858(6) & 1.796 &  $11/6$ \\
      5  &  2.92254 & 2.922318(6) & 2.113 &  2.080(7)\\
      6  &  3.17838 & 3.178520(4) & 2.232 &  \phantom{1}2.261(12)\\
      7  &  3.38403 & 3.384080(5) & 2.357 & 2.40(2)\phantom{1} \\
      8  &  3.55484 & 3.554827(4) & 2.441 & $5/2$ \\
      9  &  3.70057 & 3.700523(10) & 2.489 & $5/2$  
  \end{tabular}
  \caption{Growth rates $\lambda_d$ and exponents $\theta_d$ obtained
    from extrapolating the enumeration data. The columns marked MC contain values from large scale
    Monte Carlo simulations \cite{hsu:nadler:grassberger:05,hsu:nadler:grassberger:05a}.}
  \label{tab:lambda}
\end{table}

The same approach can be used to compute the exponent $\Theta_d$. Here
we expect
\begin{equation}
  \label{eq:theta-finite-size}
  \Theta_d(n) \sim \Theta_d + \frac{b}{n^{\Delta}}\,.
\end{equation}
Again we used the data points $\Theta_d(n)$ for the three largest
values of $n$ to fit the parameters $\Theta_d$, $b$ and $\Delta$. 
Figure~\ref{fig:theta} shows that $\Theta_d(n)$ in fact scales like
\eqref{eq:theta-finite-size}. The resulting estimates for $\Theta_d$
(Table~\ref{tab:lambda}) deviate from the Monte Carlo results and the 
exact values by no more than 3\%. 

The estimates for $\lambda_d$ and $\Theta_d$ based on the current
known values of $A_d(n)$ are much more precise than previous
extrapolations based on shorter sequences $A_d(n)$, see \cite{gaunt:sykes:ruskin:76,gaunt:80}.

\section{Conclusions}

We have seen that the memory requirements of Redelmeier's algorithm
can be kept low by using hyperspherical regions of the lattice. Even
in high dimensions, the limiting resource in Redelmeier's algorithm is
time, not space. 

We used a lean and efficient implementation of Redelmeier's algorithm
to compute new perimeter polynomials in hypercubic lattices of
dimensions $d\leq 10$. We have used these new perimeter polynomials
together with combinatorial arguments based on proper animals to
compute new values of the cluster numbers $A_d(n)$ and
new formulas for $A_d(n)$ for $n\leq 14$ and arbitrary $d$. We've also
used the new data to compute formulas for the perimeter polynomials
$P_d(n;q)$ for $n\leq 11$ and arbitrary $d$. We haven't shown these
formulas here, but you can download them from the project webpage
\cite{animals:site}.

We've also used our data to compute the formula for
$\DX(n,n-7)$, the number of proper animals of size $n$ in dimension
$n-7$, and new coefficients in the series expansion of the mean
cluster size $S(p)$. 

Based on the enumeration data,  we've finally computed numerical values for the
growth rates $\lambda_d$ and the critical exponents $\theta_d$ that
agree very well with the results of Monte Carlo simulations and field
theoretical predictions. 

All in all we have explored the limits of computerized counting of lattice animals in
dimensions $d\geq 3$. Any significant extension of the results presented here
would require either a considerable amount of CPU time or an algorithmic
breakthrough comparable to the transfer matrix methods for $d=2$. 

\appendix
\section{Formulas for $\DX(n,n-k)$: Structure}
\label{sec:dx-structure}

In the physics literature like \cite{peard:gaunt:95}, equation
\eqref{eq:dx-form} is usually \emph{assumed} to be true just because
it is supported by the available enumeration data. But as a matter of fact, one can
actually \emph{motivate} \eqref{eq:dx-form} using the type of arguments that
were used in \cite{barequet:barequet:rote:10} to prove the formula
for $\DX(n,n-2)$. The idea is to show that the
leading order of $\DX(n,n-k)$ is $\sim 2^n n^{n+k-4}$ whereas the
lowest order contributions are $\sim 2^n n^{n-2k-1}$. This is exactly
the range of terms in \eqref{eq:dx-form} if $g_k$ is a polynomial of
degree $3(k-1)$.

Equation \eqref{eq:dx-form} is obviously correct for $k=1$ (with
$g_1=1$).  For $k>1$, we can still represent lattice animals 
by trees, namely the spanning trees of their adjacency graph.
Each spanning tree is again an edge labeled tree, but this time there
are only $n-k$ labels for $n-1$ edges, i.e., $k-1$ edges will
carry a label that is also used elsewhere in the tree. Consider a tree
whose $n-1$ edges are labeled with numbers $1,\ldots,n-1$. If we
identify each of the high value labels $n-k+1, n-k+2,\ldots,n-1$ with
one of the low value labels $1,\ldots,n-k$, we get the right set of
labels. Since there are $(n-k)^{k-1}$ ways to this, the number of edge
labeled trees with $n-k$ distinct labels scales like $n^{k-1} n^{n-3}
= n^{n+k-4}$ in leading order. With two directions for every edge we
get $2^n n^{n+k-4}$ for the leading order in $\DX(n,n-k)$.

\begin{figure}
  \centering
    \psset{unit=0.1\linewidth}
    \begin{pspicture}(-0.2,-0.4)(4.2,1.4)
      \graphnodes
      \Cnode(0,0){lu}
      \Cnode(1,0){ru}
      \Cnode(0,1){lo}
      \Cnode(1,1){ro}

      \psset{shadow=false,fillstyle=none}
      \ncline{->}{lu}{ru} \nbput{$i$}
      \ncline{<->}{ru}{ro} \nbput{$j$}
      \ncline{<-}{ro}{lo} \nbput{$i$}
      \ncline[linestyle=dashed]{-}{lo}{lu}

      \pscurve(2.6,1.0)(3.2,0.8)(3.0,1.4)
      \pscurve(4.4,1.0)(3.8,0.8)(4.0,1.4)
      \pscurve(2.6,0.0)(3.2,0.2)(3.0,-0.4)
      \pscurve(4.4,0.0)(3.8,0.2)(4.0,-0.4)
      \graphnodes
      \Cnode(3,0){lu}
      \Cnode(4,0){ru}
      \Cnode(3,1){lo}
      \Cnode(4,1){ro}
   \end{pspicture}
  \caption{A labeled spanning tree that contains a part
    $\Circle\stackrel{i}{\longrightarrow}\Circle\stackrel{j}{\longleftrightarrow}\Circle\stackrel{i}{\longleftarrow}\Circle$
    corresponds to a $4$-loop in the lattice animals, a quadrilateral
    that lives in the $i$-$j$ plane (left). To count the number of spanning trees
    with such a $4$-loop, the edges of the quadrilateral are removed
    and the vertices of the quadrilateral are considered the root
    vertices of disconnected trees (right). The number of the latter
    is given by \eqref{eq:forest-count} with $\ell=4$.
  \label{fig:tree-counting-4}}
\end{figure}

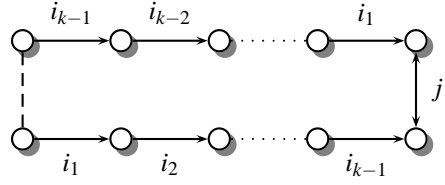
\begin{figure}
  \centering
    \psset{unit=0.1\linewidth}
    \begin{pspicture}(-0.2,-0.4)(4.2,1.4)
      \graphnodes
      \Cnode(0,0){lu}
      \Cnode(1,0){mu}
      \Cnode(2,0){nu}     
      \Cnode(3,0){ou}
      \Cnode(4,0){ru}
       \Cnode(0,1){lo}
      \Cnode(1,1){mo}
      \Cnode(2,1){no}     
      \Cnode(3,1){oo}
      \Cnode(4,1){ro}

      \psset{shadow=false,fillstyle=none}
      \ncline{->}{lu}{mu} \nbput{$i_1$}
      \ncline{->}{mu}{nu} \nbput{$i_2$}
       \ncline[linestyle=dotted]{-}{nu}{ou} 
      \ncline{->}{ou}{ru} \nbput{$i_{k-1}$}
      \ncline{<->}{ru}{ro} \nbput{$j$} 
      \ncline{<-}{ro}{oo} \nbput{$i_1$}
      \ncline[linestyle=dotted]{-}{oo}{no} 
      \ncline{<-}{no}{mo} \nbput{$i_{k-2}$}
      \ncline{<-}{mo}{lo} \nbput{$i_{k-1}$}
      \ncline[linestyle=dashed]{lo}{lu}
  \end{pspicture}
  \caption{An animal that contributes to
    $\DX(n,n-k)$ uses at most $k-1$ dimensions more than once, and the
    longest loop arises when each of these $k-1$ dimensions is
    explored twice. Such a loop contains $2k$ vertices, as shown here. 
  \label{fig:tree-counting-k}}
\end{figure}

For $k>1$, a proper animal can contain
loops. Figure~\ref{fig:tree-counting-4} shows the simplest case: a
loop that arises because one dimension ($i$) is explored twice. The
result is a quadrilateral that lives in the $i$-$j$ plane of the
lattice.  In the spanning tree, this corresponds to a part that is
labeled
$\Circle\stackrel{i}{\longrightarrow}\Circle\stackrel{j}{\longleftrightarrow}\Circle\stackrel{i}{\longleftarrow}\Circle$.
Since graphs with loops have several spanning trees, our count of edge
labeled trees overcounts the number of proper animals.  We need to
subtract some contributions from loopy animals.  The idea is to break
up the part of the spanning tree that corresponds to the loop and to
separately count the number of trees that are attached to the vertices
on the loopy part. 

Consider an animal with a loop that contains $\ell$ cells and one of
its spanning trees. If we remove all the edges from the spanning tree
that connect the vertices in the loop, the remaining graph is a
forest, i.e., a collection of trees, where each tree is rooted in one
of the $\ell$ vertices.  The forest has $n-\ell$ vertices with a
total of $n-\ell$ edges, and it is independent from the way that the
root vertices have been connected in the loop.

Now the number of ordered sequences of $\ell \geq 1$ directed rooted
trees with a total of $n-\ell$ edges and $n-\ell$ distinct edge labels
is
\begin{equation}
  \label{eq:forest-count}
  2^{n-\ell} n^{n-\ell-1}\ell\,.
\end{equation}
See \cite{barequet:barequet:rote:10} for a proof of
\eqref{eq:forest-count}. The lowest order corrections come from
those animals for which the number $\ell$ of cells in a loop
is maximal.  This is the case for a loop that joins all $k-1$
non-unique edge labels, see Figure~\ref{fig:tree-counting-k}.
The number of vertices in these loops is $\ell=2k$, hence the
lowest order corrections are $\sim 2^{n-2k}n^{n-2k-1}$. 

If we want the exact number of spanning trees for loopy animals, we
need to count the number of ways to reconnect the roots of the forest
to form a single tree. But since this number does depend on $k$
but not on $n$ and we are interested only in the scaling with $n$, we
don't need to enter this discussion here. The same is true for animals
that contain several small loops instead of a single loop of maximal
length. If we apply the separation trick to one of the shorter loops,
we get a scaling of order larger than  $\sim 2^{n-2k}n^{n-2k-1}$, and
the resulting forest is then labeled with fewer labels than edges,
which increases the order even further. So the lowest order
corrections from loops come in fact from single loops of maximum
length, and these contributions are of order $\sim
2^{n-2k}n^{n-2k-1}$, as claimed in \eqref{eq:dx-form}.  

Besides the non-uniquess of spanning trees for loopy graphs, there is
another type of error that needs to be corrected: some edge labeled
trees correspond to animals with overlapping cells, i.e., to illegal
animals. For instance, if a spanning tree contains the subtree
\begin{displaymath}
  \CIRCLE\stackrel{i}{\longrightarrow}\Circle\stackrel{i}{\longleftarrow}\CIRCLE
  \quad\mbox{or}\quad
  \CIRCLE\stackrel{i}{\longleftarrow}\Circle\stackrel{i}{\longrightarrow}\CIRCLE\,,
\end{displaymath}
the two $\CIRCLE$'s represent the very same cell of the animal. But
these ``colliding'' configurations can be interpreted as $3$-loops, or
more generally, as $\ell$ loops, and counted in the same way as the
legal loops above. Again the lowest order contributions come from the
longest ``colliding'' loops which are formed by $k-1$ labels assigned
to two edges each and arranged like
\begin{displaymath}
  \CIRCLE\stackrel{i_1}{\longrightarrow}\Circle\stackrel{i_2}{\longrightarrow}
   \Circle \cdots
   \Circle\stackrel{i_{k-1}}{\longrightarrow}\Circle\stackrel{i_{1}}{\longleftarrow}\Circle
   \stackrel{i_{2}}{\longleftarrow}\Circle\cdots\Circle\stackrel{i_{k-1}}{\longleftarrow}\CIRCLE\,.
\end{displaymath}
These longest collision loops contain $2k-1$ vertices of the tree
(representing $2k-2$ cells of the animal). According to \eqref{eq:forest-count}, their number scales
like $2^{n-2k+1}n^{n-2k}$, one order above the lowest order of legal
loops.

This concludes the motivation of \eqref{eq:dx-form}. Note that a \emph{proof}
of \eqref{eq:dx-form} would require a thorough analysis to exclude contributions outside the 
range covered by \eqref{eq:dx-form}.

\section{Formulas for $\DX(n,n-k)$: Coefficients}
\label{sec:dx-coeff}

Having established the fact that $\DX(n,n-k)$ is given by
\eqref{eq:dx-form} we still have to determine the coefficients of the
polynomials $g_k(n)$. Since $g_k$ has degree $3(k-1)$, it seems that
we need to know the $3k-2$ values
$\DX(k,0),\DX(k+1,1),\ldots,\DX(4k-3,3k-3)$ to compute the
coefficients. In terms of our enumeration data this means knowledge of
$A_d(n)$ for $n \leq 4k-3$ and $d \leq n-k$. The data in
Table~\ref{tab:numbers} suffices to compute the coefficients of $g_k$
for $k=2,3$, but not for $k\geq 4$. Nevertheless we can compute 
$g_k$ for $k \leq 7$ by \emph{assuming} that the ``free energy''
\begin{displaymath}
  \lim_{n\to\infty} \frac{1}{n}\,\log A_d(n)
\end{displaymath}
has a well defined $1/d$ series expansion. This approach has been used
to compute $g_k$ for $k \leq 6$ from much less enumeration data in
\cite{peard:gaunt:95} and \cite{gaunt:peard:00}, and we used it to
compute $g_7$ from the new enumeration data.  Since the method hasn't
been described in detail elsewhere, we provide a description in this
Appendix.

Let's start with Lunnon's equation, which tells us
that $A_d(n)$ is a polynomial of degree $n-1$ in $d$ with coefficients
that depend on $n$. For $d \geq n$ we have
\begin{equation}
  \label{eq:Ad-poly}
  A_d(n) = \sum_{k=1}^{n-1} \DX(n,k){d \choose k} = \sum_{j=1}^{n-1} a_j(n)\,d^j
\end{equation}
with
\begin{equation}
  \label{eq:def-aj}
  a_j(n) = \sum_{k=1}^{n-1} \frac{\DX(n,k)}{k!}\,\left[{k \atop j}\right]\,,
\end{equation}
where $\left[{k\atop j}\right]$ denotes the Stirling number of the
first kind.  In particular we get
\begin{eqnarray*}
  a_{n-1}(n) &=& \frac{\DX(n,n-1)}{(n-1)!} \\
  a_{n-2}(n) &=& \frac{\DX(n,n-2)}{(n-2)!}+\frac{\DX(n,n-1)}{(n-1)!} \left[{{n-1}\atop{n-2}}\right]\\
 a_{n-3}(n) &=& \frac{\DX(n,n-3)}{(n-3)!}+\frac{\DX(n,n-2)}{(n-2)!}\left[{{n-2}\atop{n-3}}\right]+\frac{\DX(n,n-1)}{(n-1)!}\left[{{n-1}\atop{n-3}}\right] \\
\end{eqnarray*}
and so on. 
From \eqref{eq:Ad-poly} we get
\begin{displaymath}
  A_d(n) = a_{n-1}(n) d^{n-1} \left(1 +
    \sum_{j=1}^{n-2}\frac{a_{n-1-j}}{a_{n-1}} d^{-j}\right)\,,
\end{displaymath}
and with $\ln(1+x)=x+x^2/2+x^3/3\ldots$ this gives the $1/d$ series
for the ``free energy'' 
\begin{equation}
  \label{eq:ln-Ad-series}
  \frac{1}{n}\ln A_d(n) = \left(1-\frac{1}{n}\right)\ln d +
  \frac{1}{n}\ln a_{n-1}(n) + \frac{1}{n}\,\frac{a_{n-2}(n)}{
    a_{n-1}(n)}\,\frac{1}{d} + \bigo{\frac{1}{d^2}}\,.
\end{equation}
We \emph{assume} that all coefficients in this series remain bounded in
the limit $n\to\infty$. This is definitely true for the zeroth order
term:
\begin{displaymath}
  \lim_{n\to\infty}\frac{1}{n}\ln a_{n-1}(n) = 1 + \ln 2\,.
\end{displaymath}
For the first order coefficient we get
\begin{eqnarray*}
  \frac{1}{n} \frac{a_{n-2}}{a_{n-1}} &=&
  \frac{n-1}{n}\frac{\DX(n,n-2)}{\DX(n,n-1)} + \frac{1}{n} \left[{n-1
      \atop n-2}\right] \\
   &=&\left(1-\frac{1}{n}\right) \left(\frac{g_2(n)}{4n^2} - \frac{n-2}{2}\right)\,.
\end{eqnarray*}
This is only bounded if the $g_2(n)$ term balances the second term,
i.e., if the $n^3$ coefficient of $g_2$ equals $2$. 
Using also the fact that $\DX(2,0)=0$, we can write
\begin{displaymath}
  g_2(n) = (n-2)(2n^2+bn+c)\,.
\end{displaymath}
To compute the remaining coefficents, we only need to know $\DX(3,1) =
1$ and
\begin{displaymath}
  \DX(4,2) = A_2(4) - 2 = 17
\end{displaymath}
to get
\begin{displaymath}
  g_2(n) = (n-2)(2n^2-6n+9)\,.
\end{displaymath}
The postulation of bounded coefficients in
the series \eqref{eq:ln-Ad-series} has saved us from knowing the value
$\DX(5,3)$ to compute $g_2$. How much does it help us to compute $g_k$? 

The polynomial $g_k$ enters the series expansion
\eqref{eq:ln-Ad-series} via the term
\begin{displaymath}
 \frac{1}{n} \frac{\DX(n,n-k)}{\DX(n,n-1)}\frac{(n-1)!}{(n-k)!} =
  2^{2-2k} n^{1-2k} g_k(n) \underbrace{(n-1)(n-2)\cdots(n-k+1)}_{\Theta(n^{k-1})}
\end{displaymath}
in the coefficient of $d^{-(k-1)}$. The leading order of this term is
$n^{-k} g_k(n)$. All terms of degree larger than $k$ in the polynomial
$g_k$ lead to unbounded contributions to the series coefficient that
need to be counterbalanced by other terms. These balancing terms
always exist, a fact that gives additional support for the claim of
bounded coefficients. The coefficients of the terms of order larger
than $k$ in $g_k$  are therefore 
computable from the known terms $g_{k-1}(n), g_{k-2}(n),\ldots$ that
also enter the same coefficient. Only the $k+1$ low order terms of
$g_k$ are not fixed by the postulate of bounded coefficients and we
need $k+1$ data points $\DX(k,0),\DX(k+1,1),\ldots,\DX(2k,k)$ to
complete $g_k$.

Our enumeration data suffices to compute $g_7$ (see
Table~\ref{tab:gk}). The computation of $g_8$ requires knowledge
of $\DX(15,7)$ and $\DX(16,8)$, or in terms of $A_d(n)$,
\begin{displaymath}
  \DX(15,7) = A_7(15)-572521427068702741
\end{displaymath}
and
\begin{eqnarray*}
  \DX(16,8) = & A_8(16)+48366334433679758-56*A_5(16)\\
                      &+28*A_6(16)-8*A_7(16)\,.
\end{eqnarray*}


\section*{References}

\bibliographystyle{unsrt} 
\bibliography{animals,math,mertens}

\begin{thebibliography}{10}

\bibitem{golomb:book}
Solomon~W. Golomb.
\newblock {\em Polyominoes}.
\newblock Princeton University Press, Princeton, New Jersey, 2nd edition, 1994.

\bibitem{stauffer:aharony:book}
Dietrich Stauffer and Amnon Aharony.
\newblock {\em Introduction to Percolation Theory}.
\newblock Taylor~\&~Francis, 2nd edition, 1992.

\bibitem{guttmann:LNP}
Anthony~J. Guttmann, editor.
\newblock {\em Polygons, Polyominoes and Polycubes}, volume 775 of {\em Lecture
  Notes in Physics}.
\newblock Springer-Verlag, Heidelberg, 2009.

\bibitem{klarner:67}
David~A. Klarner.
\newblock Cell growth problems.
\newblock {\em Canadian Journal of Mathematics}, 19:851--863, 1967.

\bibitem{madras:99}
Neal Madras.
\newblock A pattern theorem for lattice clusters.
\newblock {\em Annals of Combinatorics}, 3:357--384, 1999.

\bibitem{barequet:barequet:rote:10}
Ronnie Barequet, Gill Barequet, and G\"unther Rote.
\newblock Formulae and growth rates of high-dimensional polycubes.
\newblock {\em Combinatorica}, 30(3):257--275, 2010.

\bibitem{redelmeier:81}
D.~Hugh Redelmeier.
\newblock Counting polyominoes: Yet another attack.
\newblock {\em Discrete Mathemetics}, 36(2):191--203, 1981.

\bibitem{mertens:90}
Stephan Mertens.
\newblock Lattice animals: A fast enumeration algorithm and new perimeter
  polynomials.
\newblock {\em J.~Stat.~Phys.}, 58(5/6):1095--1108, 1990.

\bibitem{mertens:lautenbacher:92}
Stephan Mertens and Markus~E. Lautenbacher.
\newblock Counting lattice animals: A parallel attack.
\newblock {\em J.~Stat.~Phys.}, 66:669--678, 1992.

\bibitem{jensen:01}
Iwan Jensen.
\newblock Enumerations of lattice animals and trees.
\newblock {\em Journal of Statistical Physics}, 102(3/4):865--881, 2001.

\bibitem{aleksandrowicz:barequet:09a}
Gadi Aleksandrowicz and Gill Barequet.
\newblock Counting $d$-dimensional polycubes and nonrectangular planar
  polyominoes.
\newblock {\em International Journal of Computational Geometry \&
  Applications}, 19(3):215--229, 2009.

\bibitem{aleksandrowicz:barequet:09b}
Gadi Aleksandrowicz and Gill Barequet.
\newblock Counting polycubes without the dimensionality curse.
\newblock {\em Discrete Mathematics}, 309:4576--4583, 2009.

\bibitem{conway:sloane:97}
J.~H. Conway and N.~J.~A. Sloane.
\newblock Low-dimensional lattices. {VII} coordination sequences.
\newblock {\em Proceedings of the Royal Society A}, 453:2369--2389, 1997.

\bibitem{gaunt:sykes:ruskin:76}
D.S. Gaunt, M.F. Sykes, and Heather Ruskin.
\newblock Percolation processes in $d$-dimensions.
\newblock {\em Journal of Physics A: Mathematical and General},
  9(11):1899--1911, 1976.

\bibitem{animals:site}
Stephan Mertens.
\newblock Lattice animals.
\newblock \url{http://www.ovgu.de/mertens/research/animals}, 2011.

\bibitem{gaunt:80}
D.S. Gaunt.
\newblock The critical dimension for lattice animals.
\newblock {\em Journal of Physics A: Mathematical and General}, 13:L97--L101,
  1980.

\bibitem{lunnon:75}
W.F. Lunnon.
\newblock Counting multidimensional polyominoes.
\newblock {\em The Computer Journal}, 18(4):366--367, 1975.

\bibitem{cayley:1889}
Arthur Cayley.
\newblock A theorem on trees.
\newblock {\em Quarterly Journal of Pure and Applied Mathematics}, 23:376--378,
  1889.

\bibitem{cameron:95}
Peter~J. Cameron.
\newblock Counting two-graphs related to trees.
\newblock {\em Electronic Journal of Combinatorics}, 2:R4, 1995.

\bibitem{fisher:essam:61}
Michael~E. Fisher and John~W. Essam.
\newblock Some cluster size and percolation problems.
\newblock {\em Journal of Mathematical Physics}, 2:609--619, 1961.

\bibitem{peard:gaunt:95}
P.J. Peard and D.S. Gaunt.
\newblock $1/d$-expansions for the free energy of lattice animal models of a
  self-interacting branched polymer.
\newblock {\em Journal of Physics A: Mathematical and General}, 28:6109--6124,
  1995.

\bibitem{adler:etal:88}
J.~Adler, Y.~Meir, A.B. Harris, A.~Aharony, and J.A.M.S. Duarte.
\newblock Series study of random animals in general dimensions.
\newblock {\em Physical Review B}, 38, 4941--4954 1988.

\bibitem{parisi:sourlas:81}
Giorgio Parisi and Nicolas Sourlas.
\newblock Critical behavior of branched polymers and the {L}ee-{Y}ang edge
  singularity.
\newblock {\em Physical Review Letters}, 46(14):871--874, 1981.

\bibitem{imbrie:03}
John~Z. Imbrie.
\newblock Dimensional reduction and crossover to mean-field behavior for
  branched polymers.
\newblock {\em Annales Henri Poincar\'e}, 4:S445--S458, 2003.

\bibitem{dhar:83}
Deepak Dhar.
\newblock Exact solution of a directed-site animals-enumeration problem in
  three dimensions.
\newblock {\em Physical Review Letters}, 51(10):853--856, 1983.

\bibitem{lubensky:isaacson:79}
T.~C. Lubensky and Joel Isaacson.
\newblock Statistics of lattice animals and dilute branched polymers.
\newblock {\em Physical Review A}, 20(5):2130--2146, 1979.

\bibitem{hsu:nadler:grassberger:05}
Hsiao-Ping Hsu, Walter Nadler, and Peter Grassberger.
\newblock Simulations of lattice animals and trees.
\newblock {\em Journal of Physics A: Mathematical and General}, 38:775--806,
  2005.

\bibitem{hsu:nadler:grassberger:05a}
Hsiao-Ping Hsu, Walter Nadler, and Peter Grassberger.
\newblock Statistics of lattice animals.
\newblock {\em Computer Physics Communications}, 169:114--116, 2005.

\bibitem{gaunt:peard:00}
D.S. Gaunt and P.J. Peard.
\newblock $1/d$-expansions for the free energy of weakly embedded site animal
  models of branched polymers.
\newblock {\em Journal of Physics A: Mathematical and General}, 33:7515--7539,
  2000.

\end{thebibliography}

\end{document}